\documentclass[sigconf]{acmart}

\usepackage{multirow}
\usepackage{todonotes}
\usepackage{todonotes}
\usepackage{enumitem}
\usepackage{hyperref}
\usepackage{graphicx}
\usepackage{subcaption}

\AtBeginDocument{%
  \providecommand\BibTeX{{%
    \normalfont B\kern-0.5em{\scshape i\kern-0.25em b}\kern-0.8em\TeX}}}

\copyrightyear{2023}
\acmYear{2023}
\setcopyright{acmlicensed}\acmConference[ICAIF '23]{4th ACM International Conference on AI in Finance}{November 27--29, 2023}{Brooklyn, NY, USA}
\acmBooktitle{4th ACM International Conference on AI in Finance (ICAIF '23), November 27--29, 2023, Brooklyn, NY, USA}
\acmPrice{15.00}
\acmDOI{10.1145/3604237.3626858}
\acmISBN{979-8-4007-0240-2/23/11}

\begin{document}

\title{Robust Market Making: To Quote, or not To Quote}

\author{Ziyi Wang}
\email{ziyi.1.wang@kcl.ac.uk}
%%\orcid{1234-5678-9012}
\affiliation{%
  \institution{King's College London}
  \city{London}
  \country{United Kingdom}
}

\author{Carmine Ventre}
\email{carmine.ventre@kcl.ac.uk}
\affiliation{%
  \institution{King's College London}
  \city{London}
  \country{United Kingdom}
\email{carmine.ventre@kcl.ac.uk}
}
\author{Maria Polukarov}
\email{maria.polukarov@kcl.ac.uk}
\affiliation{%
  \institution{King's College London}
  \city{London}
  \country{United Kingdom}
\email{maria.polukarov@kcl.ac.uk}
}

\renewcommand{\shortauthors}{Wang et al.}

\begin{abstract}
 Market making is a popular trading strategy, which aims to generate profit from the spread between the quotes posted at either side of the market. It has been shown that training market makers (MMs) with adversarial reinforcement learning allows to overcome the risks due to changing market conditions and to lead to robust performances. Prior work assumes, however, that MMs keep quoting throughout the trading process, but in practice this is not required, even for ``registered'' MMs (that only need to satisfy quoting ratios defined by the market rules). 
 In this paper, we build on this line of work %on robust market making, 
 and enrich the strategy space of the MM by allowing to occasionally not quote or provide single-sided quotes. 
Towards this end, in addition to the MM agents that provide continuous bid-ask quotes, we have designed two new agents with increasingly richer action spaces. The first has the option to provide bid-ask quotes or refuse to quote. %, for a two-action space. 
The second has the option to provide bid-ask quotes, refuse to quote, or only provide single-sided ask or bid quotes. %, for a four-action space. 
We employ a model-driven approach to empirically compare the performance of the continuously quoting MM with the two agents above in various types of adversarial environments. We demonstrate how occasional refusal to provide bid-ask quotes improves returns and/or Sharpe ratios. The quoting ratios of well-trained MMs can basically meet any market requirements, reaching up to 99.9$\%$ in some cases. 
\end{abstract}




%% Keywords. 
 %, machine learning}

\maketitle

\section{Introduction}
Market makers (MMs) %continuously 
quote bids and asks on two-sided markets for a particular financial instrument, providing liquidity and depth to the market and profiting from the difference in the bid-ask spread. However, it is by no means an easy strategy to execute; the complex and non-stationary nature of the market environment and the consequent market price volatility can trigger large  losses for the MM. Recent research %has focused on 
addresses the risk of model misspecification for market making; %with the authors of 
adversarial reinforcement learning (ARL) can be used to derive trading strategies that are robust to adversarial and adaptively chosen market conditions, as shown in \cite{spooner2020robust}. The model consider one market maker who maximizes profits while controlling inventory risk, and plays a stochastic zero-sum game with all the other market participants, which are represented by the adversary. The ARL-trained MM shows robust performances against epistemic uncertainty. 

The assumption made in \cite{spooner2020robust} is that the MM must always quote, even when the combination of market conditions and the state of her portfolio would suggest otherwise. Why should, for example, an MM continue to quote bids for an asset which is falling and is already ``heavy'' in her inventory? Even markets who oblige MMs to be formally recognized as such, allow a certain leeway on quoting ratios. For example, London Stock exchange requires ``registered MM'' to maintain a quote ``for at least 90\% of continuous trading during the mandatory period'' \cite{LSEDM} whilst, in accordance with MiFID II, Deutsche B\"{o}rse Cash Market requires ``registered regulated MM'' to provide quotes ``during 50 percent of daily trading hours on a monthly average'' \cite{mifid2}; interestingly, this rule ceases to apply when the market is in (appropriately defined) exceptional circumstances.

\begin{figure*}
    \centering
    \includegraphics[width=\textwidth]{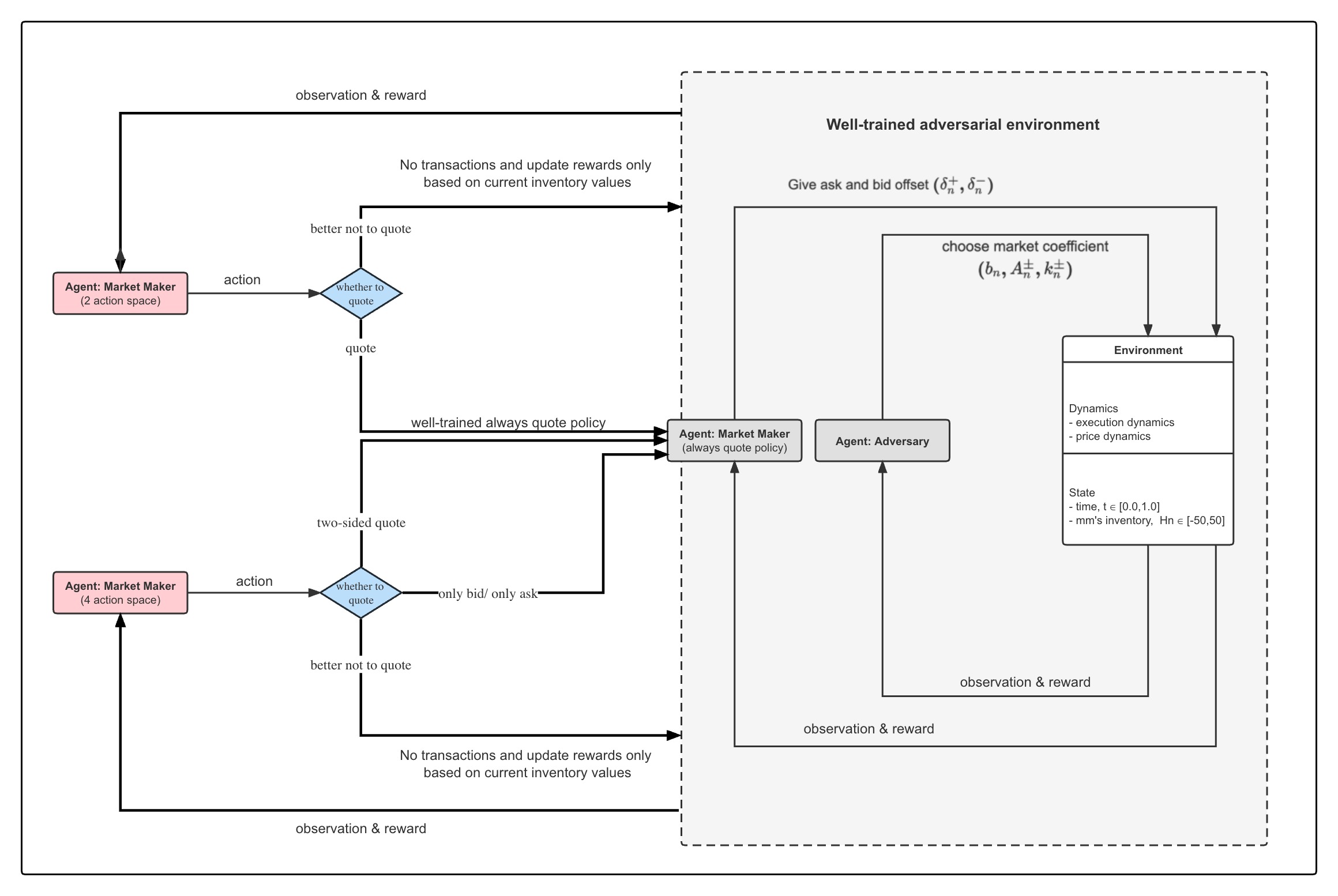}
  \caption{Training our market maker using ARL.}
  \label{fig:teaser}
\end{figure*}

In this paper, we ask whether granting major flexibility in terms of quoting ratios can improve the performances of the robust MM. %trained with ARL. 
Specifically, we add ``not to quote'' as an extra action in the MM strategy space and investigate whether the strategy can avoid some risks in the trading environment, such as excessive price volatility, and thus increase profits by not quoting. To this end, we design two new market maker agents. The former market maker has a ``2-action space'' and can choose to provide both two-sided quotes or not to quote at all. The latter has a ``4-action'' space and has the additional options of providing only bid quotes or only ask quotes. Firstly, we train a market maker agent who will always quote and give the best spread in different adversarial environments. These environments are defined by the underlying stochastic process we adopt to model the limit order book dynamics \cite{avellaneda2008high} (in other words, ours is a model-based RL approach). For this part, we follow and replicate the findings of \cite{spooner2020robust} in a different setup and with a newly developed in Python, as opposed to the one in Rust developed by \cite{spooner2020robust}, thus providing further evidence for the robustness of ARL-based market making. 
Then, we separately train market makers with two-action space and market makers with four-action space. Either can decide whether to provide two-sided quotes based on the current observed state. If they choose to quote, they will inherit the quoting ability of the agent that always quotes. In other words, they will provide an adversarially trained appropriate bid-ask spread. Otherwise, the market makers will not provide any bid or ask offset, resulting in no transactions taking place, or they may only provide bid or ask offsets, which may lead to a one-sided trade. Our overall approach is depicted in Figure \ref{fig:teaser}.

Finally, we evaluate the sensitivity of the three types of market makers %: always quoting market makers, market makers with a 2-action space, and market makers with a 4-action space, 
to three major environmental variables (namely, drift of the underlying stochastic process, order arrival rates, and order volume distribution) denoted as $B$, $A$, $K$, respectively. We also examine their performance under different risk coefficients. The results show that training market makers with adversarial reinforcement learning can effectively improve their Sharpe ratio. Additionally, allowing market makers to provide one-sided quotes and refuse quotes is also an effective means of mitigating market risks.

\section{Related Work}
Market makers profit by giving bid-ask spreads, but the imbalance between the arrival of buy and sell orders causes the market makers to accumulate unwanted inventory \cite{fodra2012high}. A maker maker is exposed to the inventory risk, which refers to the uncertainty of assets on hand due to price fluctuations \cite{guilbaud2013optimal}. There is also the execution risk, which arises from the uncertainty of when future transactions will occur \cite{ho1981optimal}. Therefore, market makers have to balance the supply and demand in the market through pricing, while actively controlling their inventory positions. Furthermore, %the authors of 
Glosten and Milgrom \cite{glosten1985bid} observe that the market maker faces an adverse selection problem, referring to the risk of information asymmetry that arises when dealing with informed traders, and try to use the role of information to explain the dynamics of price setting. This means that market makers may suffer losses when trading at the bid or ask price with informed traders who know something that the market maker does not know. But market makers can compensate for trading losses with informed traders by trading gains with liquidity traders, which also demonstrates how spreads can arise from adverse selection.

\vspace{1ex} \noindent {\bf Stochastic Control Approach.}
Many theoretical market-making models have been developed in the context of stochastic dynamic programming, such as %the works of T Ho, HR Stoll 
\cite{ho1981optimal}, where bids and offers are determined dynamically to maximize some long-term objective, such as expected profits or expected utility of profits.

%The authors of 
Ho and Stoll \cite{ho1981optimal} pioneered the study of optimal behavior of a single dealer by deriving optimal bids and offers through stochastic dynamic programming that maximizes the expected utility of a dealer's terminal wealth.  Avellaneda and Stoikov \cite{avellaneda2008high}  combined the utility framework of their method  \cite{ho1981optimal} with the statistical properties of limit order books described in the econophysics literature (cf. \cite{bouchaud2002statistical,potters2003more,smith2003statistical,luckock2003steady}). The authors proposed a two-step approach to give the best bid and ask offer. The market maker first calculates a reservation price based on her current inventory, and solves the inventory risk problem by hanging buy and sell orders around this reservation price. Second, she calibrates her bids and offers to the limit order book by considering the probability of executing her quotes as a function of their distance from the mid-price. %The authors of 
Gu\'{e}ant et al. \cite{gueant2013dealing} later solved the market making problem under inventory constraints by considering it as a stochastic control problem, similar to the one proposed in \cite{ho1981optimal} and formalized mathematically in \cite{avellaneda2008high}. Subsequently, %the authors of 
Fodra and Labadie \cite{fodra2012high} extended the  inventory-constrained market making models of  \cite{avellaneda2008high} and  \cite{gueant2013dealing} to a more general class of mid-price processes under an exponential or linear PnL utility function; they also added an inventory-risk-aversion parameter that penalizes the market makers if their inventory is not zero at the end of a day's trading. Moreover, they extended the exponential utility approach \cite{avellaneda2008high} to more general settings, including
%any Markov process with the conditional expectation that So I $E_{t,s}[S(T)]$ is affine in s, which includes 
processes like arithmetic Brownian motion drift or Ornstein-Uhlenbeck processes.

%It is important to note that 
Crucially, Avellaneda and Stoikov \cite{avellaneda2008high} assume that the MM has a perfect knowledge of the stochastic dynamics of different state variables; however,  this assumption does not correspond to reality. According to  \cite{cartea2015algorithmic}, there is no such concept as a ''correct'' model, and therefore it is not reliable to choose an optimal strategy in the wrong model. One way for MMs to address this model risk is to acknowledge that the model is misspecified. Thus, in their later work, Gu\'{e}ant \cite{gueant2017optimal} %subsequently 
investigates the impact of uncertainty in the model of \cite{avellaneda2008high}, and considers how a market maker should trade optimally, based on the assumption that the MM does not fully understands the market dynamics, i.e., by adjusting the quotes to reduce the inventory risk and adverse selection costs.

\vspace{1ex} \noindent {\bf Reinforcement Learning Approach.} 
The main limitation of these models based on  stochastic process is that to obtain a closed-form representation of the strategy, specific assumptions need to be made on, for example, the price dynamics or the order arrival process.
However, real financial markets do not conform to any simple parametric model specification with fixed parameters \cite{hambly2021recent}.
%The authors of 
To this end, Chan and Shelton \cite{chan2001electronic} were the first to %have 
explore the problem of market making using reinforcement learning. The authors tested three reinforcement learning algorithms in a simulated environment: the Monte Carlo method, the SARSA method and the actor-critic method, and found that the actor-critic method worked better than the first two %methods 
in complex environments. The stochastic strategies generated using the actor-critic approach were able to correctly adjust the bid-ask price in the presence of order imbalance and effectively control the trade-off between the profit and quoted spread. %In 
Abernethy and Kale \cite{abernethy2013adaptive} %the authors 
used an online learning algorithm to propose market maker strategies based on minimizing the quote spreads. In comparison to \cite{chan2001electronic}, Spooner et al.  \cite{spooner2018market} %subsequently 
developed a high-fidelity limit order book market simulation by using reinforcement learning to train inventory-sensitive market makers using high-frequency historical data.
However,  as mentioned above, algorithmic traders need to recognize the existence of epistemic risk, and thus, it is necessary to find a market maker strategy that can adapt to the misspecification between the training and testing environments.  Spooner and Savani \cite{spooner2020robust} were the first to apply adversarial reinforcement learning to obtain a market maker trading strategy that is robust to epistemic risk. Inspired by the ideas from robust adversarial reinforcement learning \cite{pinto2017robust}, they train the market makers %who can give optimal trading strategies 
in various %multiple 
environments, by transforming the well-studied single-agent model of \cite{avellaneda2008high} into a discrete-time zero-sum game between a market maker and an adversary. Similarly, inspired by \cite{pinto2017robust}, Ga\v{s}perov and  Kostanj\v{c}ar \cite{gavsperov2021market} use perturbations of the adversarial agent to make the MM agent more robust in modeling uncertainty, and thus improve generalization.
Furthermore, most of the models focus on dealing with market making for a single asset in an unknown financial environment. However,  reinforcement learning algorithms can often be generalized to high-dimensional control problems in a multi-asset framework; thus, Gu\'{e}ant and Manziuk \cite{gueant2019deep} employ a numerical approach to determine the optimal bid and ask quotes for a large universe of bonds. 

In this paper, we are inspired by \cite{spooner2020robust} and use a model-based adversarial reinforcement learning approach to explore the performance of market-making strategies that can flexibly accept or reject to quote, and to evaluate whether new market-making agents can balance trading returns and risks in complex market environments. Unlike most of previous reinforcement learning-based market making studies that used data-driven limit order book models,  \cite{spooner2020robust} propose to employ reinforcement learning with market dynamics modeled and explained by the optimal control approach, which allows the features of adversarial training to be examined through single-stage analysis, while minimizing systematic errors caused by frequent biases in the historical data. Existing research on reinforcement learning-based market making strategies assumes that market makers provide liquidity through continuous quotes, but this also exacerbates the risk of losses caused by the market makers having to trade in times of highly volatile market environments.

\section{Problem Description}
In this section, we define our model setting, by specifying the market dynamics and the types of behaviors for the environment (adversary) when the market maker quotes. 
%\subsection{Market Dynamics}

\vspace{1ex}\noindent{\em Market Dynamics.}
Our model is inspired by the standard model of market making proposed in \cite{avellaneda2008high} and largely based on the trading model proposed in \cite{spooner2020robust}. We use such a model to explore a larger action space motivated by regulations on quoting ratios, as discussed above. 
Typically, the price of an asset is an important trade-off factor for all trading players, including market makers, when they interact with the environment.
We assume that the price of the asset is a stochastic process throughout the trading round, which can be described using Brownian motion with drift. 
If the price of an asset is known to be $Z_n$ at $t=n$, then the expectation of the price at time $t=n+1$ depends only on the drift coefficients $b_n$ and volatility coefficients $\sigma_n$ of the Brownian motion: 
\begin{equation*}
  Z_{n+1}=Z_n+b_n\Delta t+\sigma_n W_n,
\end{equation*}
where $W_n$ is an independent, normally-distributed random variable drawn from a normal distribution $N(0,\Delta t)$ with mean $0$ and variance $\Delta t$ in any finite time interval $\Delta t$ (i.e., the variance increases linearly with the length of the time interval). Therefore, according to this model, once the values of $b_n$ and $\sigma_n$ are determined, only the current price, and not the historical prices, affect the probability distribution of future prices.  We conducted our experiment with an initial price $Z_0=100$ and fixed volatility $\sigma=2$ throughout.

For an asset, the difference between the market maker's expected buy (bid) price, denoted $P_n^+$, and the current market price of the asset, is known as the bid offset, $\delta_n^+$. The difference between the market maker's expected sell (ask) price, denoted $P_n^-$, and the current market price of the asset, is known as the ask offset, $\delta_n^-$. The bid and ask offset can be expressed as:
\begin{equation*}
  \delta_n^\pm=\pm[Z_n-P_n^\pm]\ge0.
\end{equation*}
The market maker has to give a sensible bid/ask offset based on the current asset market price $Z_n$ and the current inventory, denoted $H_n$. This offset allows the MM to hedge the inventory risk, measured by a risk coefficient, $\eta$. As mentioned in the introduction, for a trained market maker that decides not to quote, we let the offset be large above a certain threshold at any such time step $n$; for simplicity, we say that $\delta_n^\pm=+\infty$ in this case.

At a given time step, we consider the probability of executing a trade between the market maker and the environment, according to a Poisson distribution, and assuming that the current inventory of the market maker can meet its trading needs. The order arrival intensities are given by:
\begin{equation*}
  \lambda_n^\pm=A_n^\pm e^{{-k}_n^\pm\delta_n^\pm}.
\end{equation*}
That is, whether an order is executed depends not only on the bid-ask offset given by the market maker, but also on two important parameters in the environment: $A_n^\pm>0$ and ${,k}_n^\pm>0$,  describing the rate of arrival of market orders and distribution of volume in the book, respectively. These two parameters can be customized by the adversary within certain boundaries in the zero-sum game.

The cumulative return of a market maker can be regarded as the sum of the cash and inventory values, called wealth, expressed as:
\begin{equation*}
  \Pi(X,H,Z)=X+HZ,
\end{equation*}
where $X$ refers to market maker's cash: 
\begin{equation*}
\begin{aligned}
  X_{n+1}&=X_n+P_n^-\Delta N_n^--P_n^+\Delta N_n^+ 
  \\&=X_n+\delta_n^-\Delta N_n^-+\delta_n^+\Delta N_n^+-Z_n\Delta H_n, 
\end{aligned}
\end{equation*}
where $N_n^+$ ($N_n^-$) denotes the cumulative number of assets bought (sold) by the market maker at bid (ask) prices until time $t=n$ and $\Delta N_n^\pm±=N_{n+1}^\pm-N_n^\pm$. Every time a market maker completes a transaction with a trading volume of 1 unit, the profit earned is equal to offset, and the actual cash inflow or outflow is equal to bid/ask price. Also, as prices change, the value of a unit of inventory assets changes.   The current inventory quantity, $H_n$, is given by the difference between these two terms:
\begin{equation*}
  H_n=({N_n^+-N}_n^-)\in[\underline{H},\overline{H}].
\end{equation*}

\vspace{1ex}\noindent{\em Adversarial Strategies for ``Always Quote''.} 
The adversary has three core parameters of the market model dynamics that affect the prices and execution: $b_n$, $A_n^\pm$ and $k_n^\pm$, where $b_n$ is the drift coefficient, $A_n^\pm$ describes the rate of the market orders arrival, and $k_n^\pm$ describes the distribution of volume in the book.

\begin{description}[leftmargin=10pt]
\item[Fixed Adversary.] The three core parameters of the environment (adversary) remain fixed for all episodes, with $b_n=0$, $A_n^\pm=140$ and $k_n^\pm=1.5$. Thus, the  interaction process between the market maker and the environment can be regarded as a single-agent problem with stationary transition dynamics. 
\item[Random Adversary.] At the beginning of each episode, the adversary's core parameters are chosen independently and uniformly at random from the following ranges: $b_n=b\in[-5,5]$, $A_n^\pm=A\in[105,175]$, $k_n^\pm=k\in[1.125,1.875]$, and then remain fixed until the terminal stage. That is, after the realization of these three variables, the environment can be thought of as a fixed adversary, with the small difference though that the interaction with the market maker can now be viewed as a single-agent problem with non-stationary transition dynamics.
\item[Strategic Adversary.] In each episode, the adversary chooses the market coefficients $b_n$, $A_n^\pm$, $k_n^\pm$, which are bounded as in the previous setting. To model the adversary and the market maker in a zero-sum game, we set the adversary's reward to be the opposite of the market maker's. A strategic adversary will then choose the market coefficients to maximize its own reward, which is equivalent to minimizing the market maker's reward. %as well at each time step. 
In addition, to further explore the impact of each of these three variables on the trading environment, we also set up three other types of strategic adversaries, each controlling only one of the parameters $b_n$, $A_n^\pm$, or $k_n^\pm$,  respectively.
%which are strategic adversary only controlling $b_n$, strategic adversary only controlling $A_n^\pm$, and strategic adversary only controlling $k_n^\pm$ respectively.
\end{description}

\section{Trading agent based on ARL}
Here, we define our experimental setting by specifying the training process for both the adversary agents and market-maker agents. First, we train the adversary agents and the always-quoting market-maker agents (although not simultaneously) using the SAC algorithm. Then, we train market-maker agents with two action spaces and four action spaces using DQN, enriching the action space of the market-making strategy to explore how market-makers can use flexible options such as bilateral quotes, unilateral quotes, and no quotes to combat market risks.

\subsection{Training MM and adversary via SAC}
We model a discrete-time zero-sum game between a market maker and an adversary. We set the market maker's objective to maximize the expected returns while controlling the inventory and quoted spreads. The adversary's objective, on the other hand, is to suppress the MM's objective as much as possible. Thus, the adversary's reward function is set as the negative of that of the market maker. %equation to be the opposite of the market maker's. 

We use the SAC algorithm, a soft actor-critic method by \cite{haarnoja2018soft}, to train both the adversary agent and the market maker agent. We need to train the adversary first, then we train the market maker to play against the strategic adversary. The well-trained adversary strategy is stored as an attribute in the environment. At each step, the environment uses the adversarial strategy to update the market coefficients, such as the coefficient $b_n$ that affects the market price change, and market makers are required to give the bid and offer offsets, $\delta_n^\pm$. Subsequently, the Poisson distribution process that determines whether a trade occurs, is jointly influenced by the market coefficients $A_n^\pm$, $k_n^\pm$, and the market maker's offsets. Finally, new rewards and observations will be returned.

\vspace{1ex}\noindent{\em Training the MM.} The definition of the environment follows.
\begin{description}[leftmargin=10pt]
\item[State.] The state, $s_n=(t_n,H_n)$, is given by a vector composed of the current time, $t_n=\frac{nT}{N}=n\Delta t$, and the agent's inventory, $H_n$, where $t\in[0.0,\ 1.0]$ with increment $\Delta t=0.005$, and the initial inventory $H_0\in[H=-50,H=50]$.
\item[Action.] The action, $a=(\delta_n^+,\delta_n^-)$, where $\delta_n^\pm\in[-3,3]$ specify the market marker's bid and ask offsets range.
\item[Reward.] For the market maker, if a  transaction occurs, the direct reward of taking action $a$ at state $s$ is offset $\delta_n^\pm\geq0$. When market maker is buying high or selling low, the value of their inventory can vary depending on the price (see the wealth's definition above, $\Pi_n=\Pi(X_n,H_n,Z_n)$ ). 
Following  \cite{spooner2020robust}, the reward function is then given by:
\begin{equation*} 
    R_n=\Delta \Pi_n - \zeta H_n^2 -\left\{
\begin{aligned}
    &0 &for\, t<T&, \\&\eta H_n^2 &otherwise. &
\end{aligned}
\right.
\end{equation*} 
Our goal is to design a market making strategy that maximizes the wealth. If $\eta=0$ and $\zeta=0$, we have $R_n=\Delta\Pi_n$ and consider this case as risk-neutral (RN). Otherwise, the agent's reward can vary depending on the current inventory $H_n$. We refer to this case as risk-averse (RA) and study six different sets of RA coefficients in our experiment.
\item[The state transition.] We assume that the market maker trades only one unit at a time if the trading conditions are satisfactory. At each state, if the market maker gives a bid price, the market maker's inventory will increase by one unit after the execution. Similarly, the inventory will decrease by one unit after the market maker executes the trade at the ask price.
\item[Parameter configuration.] The market-making agent is trained for 30,000 episodes. The policies update every 1000 time-steps with the learning rate of $3\times{10}^{-4}$ for both the actor and critic policies. The batch size is $64$. 
\end{description}

%\vspace{1ex}
\noindent{\em Training the adversary.} 
Although the adversary and the market maker are not trained simultaneously, we set %the adversary and the market maker 
both agents with the same state space, state transition, and parameter configuration for the experiment. The difference between the two agents is in their action spaces and reward functions. In addition, to fairly compare the performance of market makers, we used the same actor and critic neural network hyperparameters to train the market makers and adversaries in all 42 sets of environments.
\begin{description}[leftmargin=10pt]
\item[Action.] For the adversaries controlling only $b_n$, only $A_n^\pm$, or only $k_n^\pm$, their action is a one-dimensional vector that corresponds to the range of values of their controlled variable. For a strategic adversary that controls all three parameters simultaneously, action ${a=(b}_n,A_n^\pm,k_n^\pm)$ is given by a vector that contains the drift, scale and decay, where $b_n=b\in[-5,5]$, $A_n^\pm=A\in[105,175]$ and $k_n^\pm=k\in[1.125,1.875]$.
\item[Reward.] To model the adversary and the market maker in a zero-sum game, we set the adversary's reward to be the opposite of that of the market maker:
\begin{equation*} 
    R_n=-\Delta \Pi_n + \zeta H_n^2 +\left\{
\begin{aligned}
    &0 &for\, t<T&, \\&\eta H_n^2 &otherwise&,
\end{aligned}
\right.
\end{equation*} 
A good adversary will choose the appropriate market coefficients to maximize its own reward, which is equivalent to minimizing the market maker's reward. %The adversary's reward function is given by:
\end{description}

\begin{figure*}
        \centering
        \includegraphics[scale=0.5]{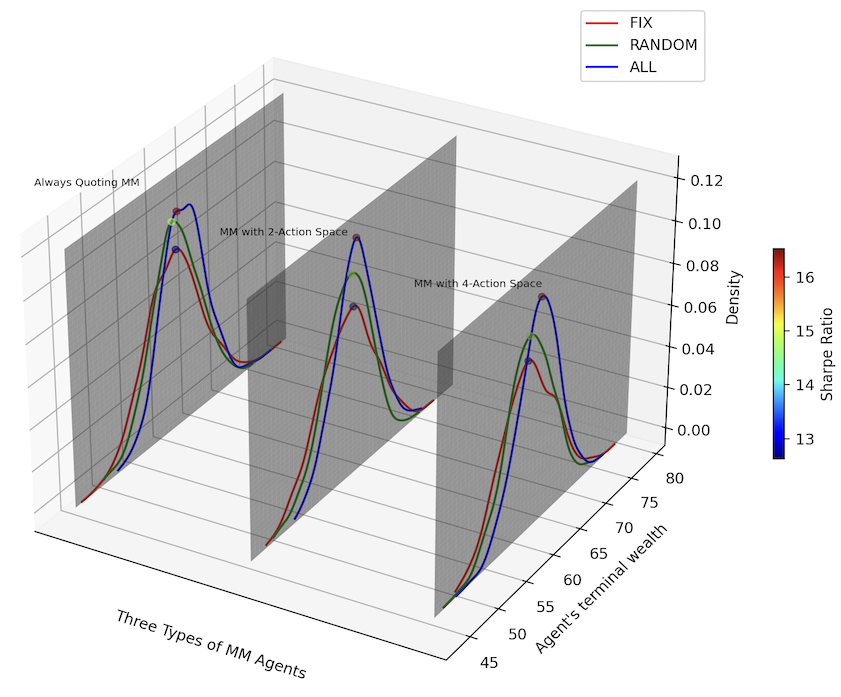}
        \caption{Performance of Three Types of MM Agents (\small $\eta=0.1, \zeta=0.0$)} \label{fig:a}
    \end{figure*}    

\begin{figure*}

    \begin{subfigure}{0.5\textwidth}
    \centering    
        \includegraphics[scale=0.4]{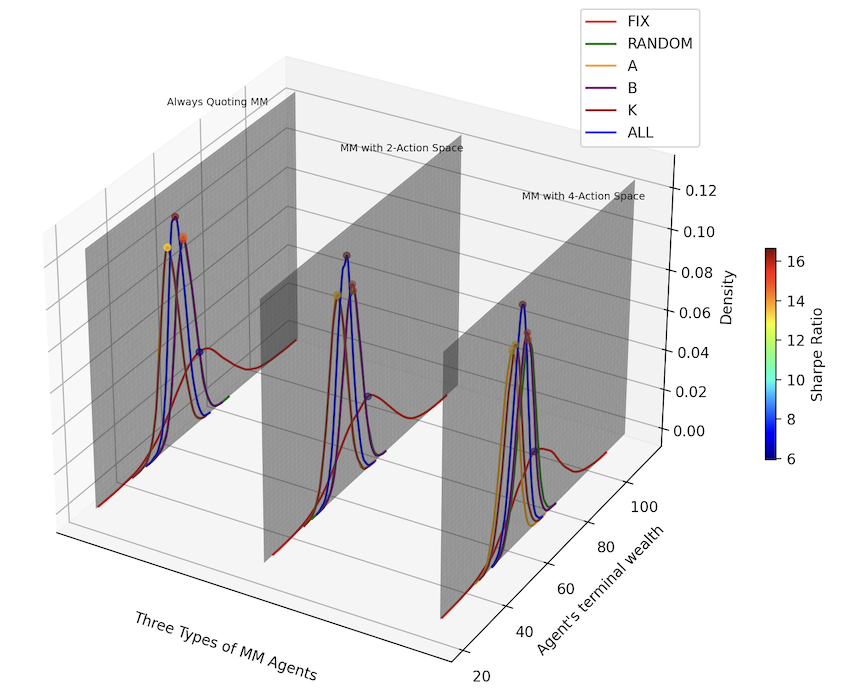}
        \caption{\small $\eta=0.0, \zeta=0.0$} \label{fig:b}
    \end{subfigure}\hspace*{\fill}
    \begin{subfigure}{0.5\textwidth}
    \centering
        \includegraphics[scale=0.4]{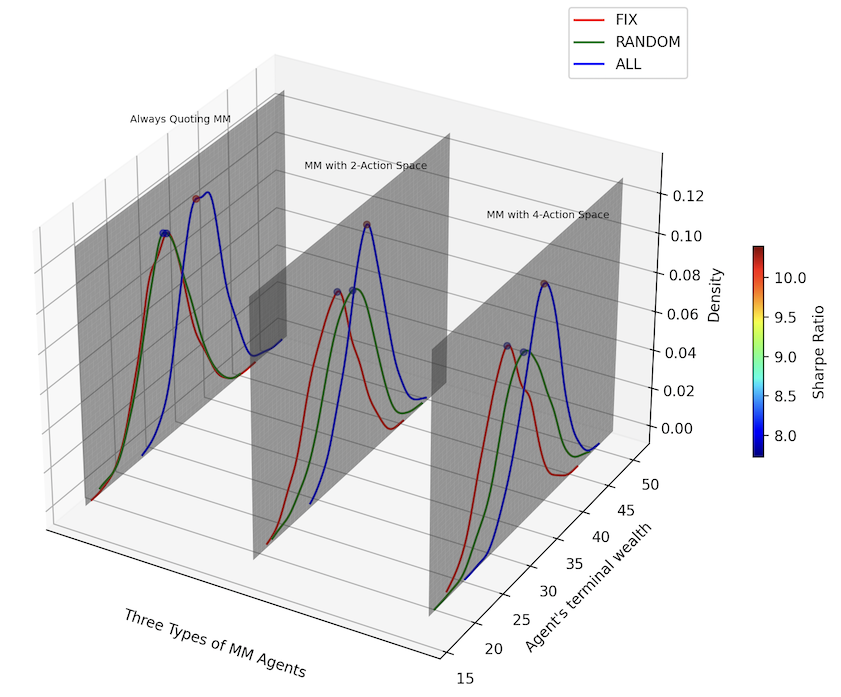}
        \caption{\small $\eta=1.0, \zeta=0.0$} \label{fig:c}
    \end{subfigure} 

    \medskip
    
    \begin{subfigure}{0.5\textwidth}
        \centering
        \includegraphics[scale=0.4]{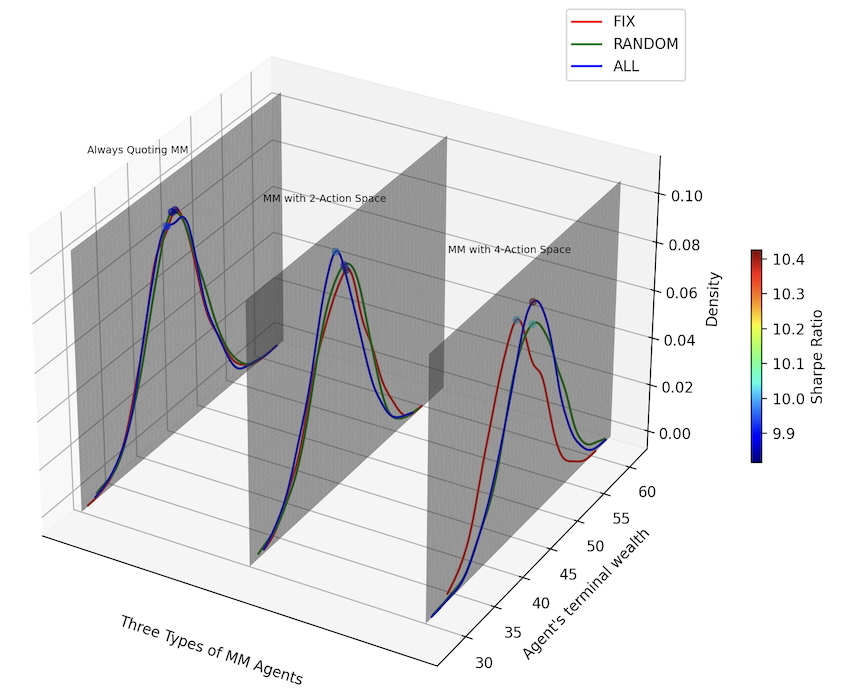}
        \caption{\small $\eta=0.5, \zeta=0.0$} \label{fig:d}
    \end{subfigure}\hspace*{\fill}
    \begin{subfigure}{0.5\textwidth}
    \centering
        \includegraphics[scale=0.4]{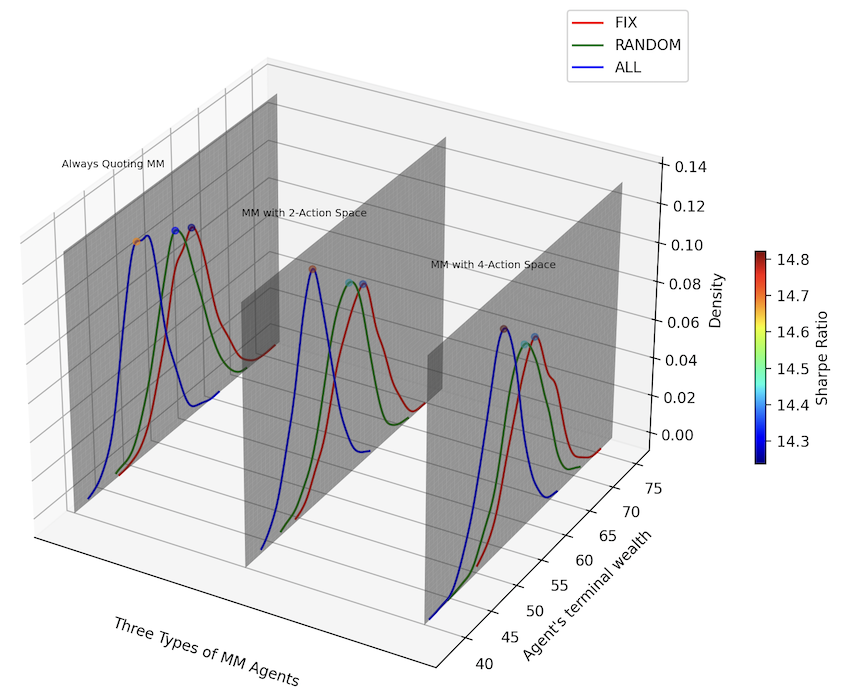}
        \caption{\small $\eta=0.01, \zeta=0.0$} \label{fig:e}
    \end{subfigure} 

    \medskip
    
    \begin{subfigure}{0.5\textwidth}
        \centering
        \includegraphics[scale=0.4]{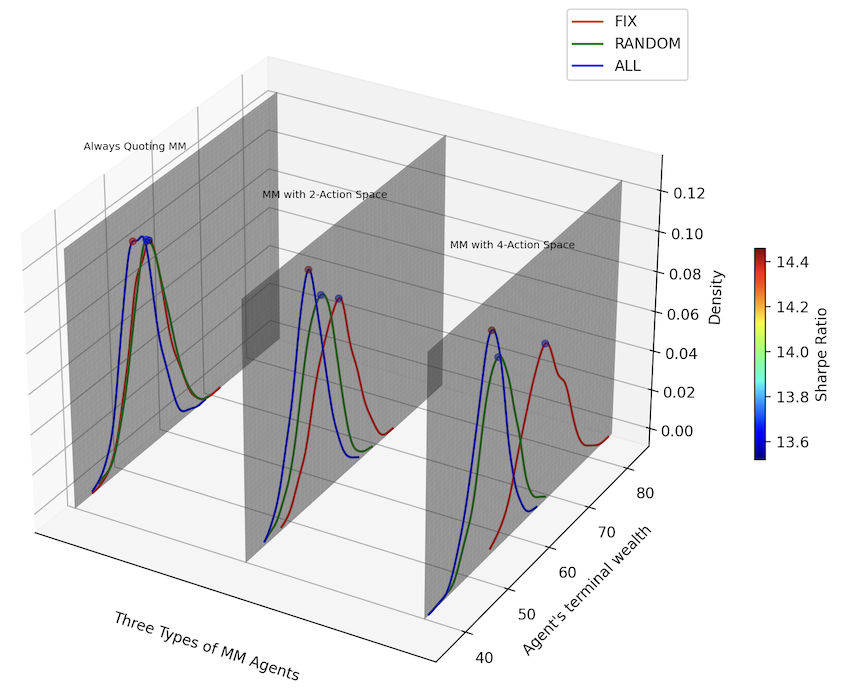}
        \caption{\small $\eta=0.0, \zeta=0.01$} \label{fig:f}
    \end{subfigure}\hspace*{\fill}
    \begin{subfigure}{0.5\textwidth}
    \centering
        \includegraphics[scale=0.4]{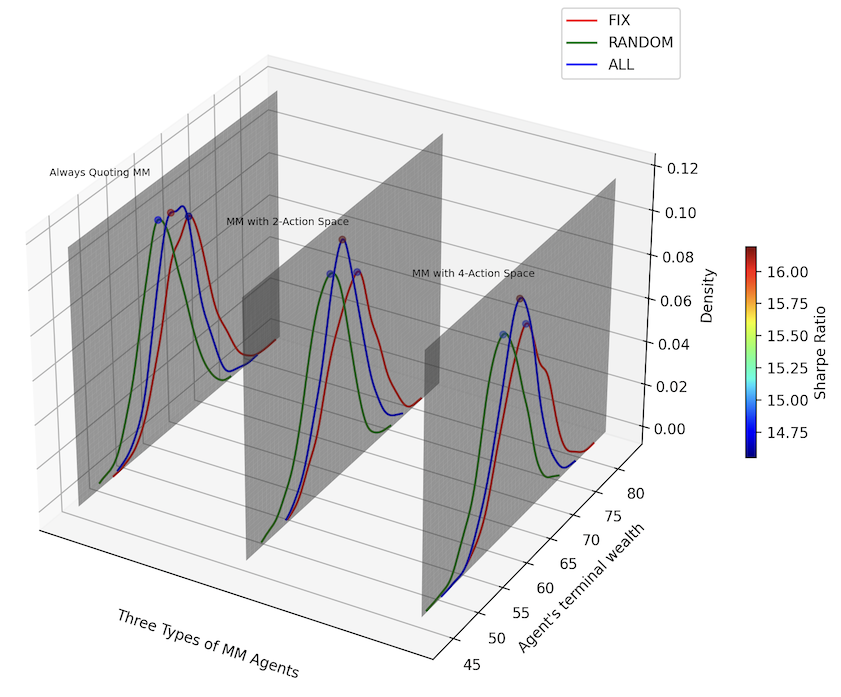}
        \caption{\small $\eta=0.0, \zeta=0.001$} \label{fig:g}
    \end{subfigure} 

\caption{Performance of Three Types of MM Agents.} \label{fig:performance}

\end{figure*}

\subsection{Training MM able to not quote via DQN}
We now have two well-trained policies, the always quoting market making policy and the adversary policy, respectively. After adding the option of not to quote, the new environment stores those two well-trained policies, where a well-trained MM policy can always give the best ask and bid offset given the observation, and a well-trained adversary policy can update the market coefficients under different situations. We then use the Deep Q Learning (DQN) agent model for the market maker to determine if it should  quote or not at all, with discrete action spaces.  

This newly designed market maker is trained in an adversarial environment, similar to the previous one, but it is able to decide whether or not to quote, based on the observation. Once it decides to quote it will give the stored bid and ask offsets. The execution dynamics, the price dynamics and the state space in the new environment are the same as before.

\begin{description}[leftmargin=10pt]
\item[State.] All environments in our experiments have the same observation specification, given by the time and the current market maker’s inventory: $s_n=(t_n,H_n)$.
\item[Action.] After adding the option not to quote, the market maker's action specification is binary, with $0$ indicating that the market maker is not quoting and $1$ indicating that it is more beneficial for the market maker to give a reasonable quote to maximize her returns in the current market conditions.
\item[Reward.] To compare whether the addition of the no-quote possibility is beneficial to improving the robustness of market makers, we set the reward function to be the same, regardless of whether the market maker always quotes or is allowed not to quote.
\item[State transition.] If the MM chooses to quote, she needs to give the bid and ask spread respectively, and market maker's inventory will increase or decrease by one unit after the execution. If she chooses not to quote, only the  %innovation of the market will happen. 
market conditions will change. Since there is no transaction execution, the market maker's inventory will naturally stay the same. %not change.
\item[Parameter configuration.] The new MM agent is trained for 30,000 episodes, and the policies update every 1 time-step with the learning rate of $1\times{10}^{-4}$. The batch size is $64$. 
\end{description}

\section{Evaluation}
Three market maker agents, including always-quoting, two-action, and four-action MM agents, are assessed for performance in standard requirements through 100 evaluations, each with 1000 episodes, in a fixed adversarial environment, calculating mean and sample standard deviation. We experimentally compare the profits and Sharpe ratios for each strategy. The Sharpe Ratio, defined as $\frac{E(\prod_N)}{\sigma(\prod_N)}$, measures the reward per unit of risk. Technically speaking, a good trading strategy yields wealth that has a larger mean and a smaller standard deviation. Figures \ref{fig:a} and \ref{fig:performance} show the performance of three agents in terms of profits under different risk environments. In these figures, each gray plane is a 2D plot with three curves representing returns distribution for different agents in various adversarial environments. The x-axis represents terminal return values, while the y-axis shows distribution intensity. The color-coded point on each curve shows its Sharpe ratio, according to the accompanying color bar. Three parallel gray planes represent agents with distinct action spaces. The use of a 3D format aids visual comparison but this additional dimension does not carry practical significance.

Since a market maker's objective is not only to maximize profits but also to effectively control the inventory, the terminal inventory of a market maker will also be used as a useful metric. We use the risk coefficient $\eta$ in the reward equation to reflect the impact of the terminal inventory on the market maker returns. However, this is not a strict requirement in this experiment. For example, in the risk-averse case ($\zeta=0$ and $\eta=0$), we assume that the market maker's wealth is unaffected by the terminal inventory.

The quoting ratio is another important metric used in this experiment, following industry standards. The market maker might need her quoting ratio to meet the requirements of the market rules. 

Tables \ref{Always Quoting MM against Adversarial Training} - \ref{MM with 4-Action Space against Adversarial Training} show the evaluation results of the three market makers in a fixed adversarial environment. The quoting ratio in the tables are presented in the format of 'no-quote ratio + bilateral quote ratio + ask-only ratio + bid-only ratio'. Additionally, we record the spread provided by the agents when they provided two-sided quotes, which is the sum of bid offset and ask offset, and calculated the mean and standard deviation of these spreads. Generally, tighter spreads imply more efficient markets.

% \usepackage{multirow}
% \usepackage
\begin{table*}[h]
\centering
\caption{Always Quoting MM against Adversarial Training}
\label{Always Quoting MM against Adversarial Training}
\resizebox{\textwidth}{!}{%
\begin{tabular}{|c|c|c|c|c|c|c|c|}
\hline
Risk   Coefficient & Desiderata & Fix & Random & A & B & K & All \\ \hline
\multirow{4}{*}{$\eta$=0.0,   $\zeta$ =0.0} & Return & 66.77±11.26 & 61.00±4.12 & 54.57±4.09 & 61.10±4.07 & 54.55±4.07 & 57.74±3.49 \\ \cline{2-8} 
 & Sharpe Ratio & 5.93 & 14.79 & 13.35 & 15.01 & 13.39 & 16.54 \\ \cline{2-8} 
 & Quoting Ratio & 0.00+100.00+0.00+0.00 & 0.00+100.00+0.00+0.00 & 0.00+100.00+0.00+0.00 & 0.00+100.00+0.00+0.00 & 0.00+100.00+0.00+0.00 & 0.00+100.00+0.00+0.00 \\ \cline{2-8} 
 & Spread & 1.68±0.26 & 1.49±0.24 & 1.80±0.34 & 1.50±0.25 & 1.72±0.31 & 1.60±0.28 \\ \hline
\multirow{4}{*}{$\eta$=1.0,   $\zeta$=0.0} & Return & 29.97±3.88 & 30.01±3.84 & - & - & - & 36.10±3.48 \\ \cline{2-8} 
 & Sharpe Ratio & 7.72 & 7.81 & - & - & - & 10.37 \\ \cline{2-8} 
 & Quoting Ratio & 0.00+100.00+0.00+0.00 & 0.00+100.00+0.00+0.00 & - & - & - & 0.00+100.00+0.00+0.00 \\ \cline{2-8} 
 & Spread & 2.30±0.95 & 2.34±1.08 & - & - & - & 2.80±0.62 \\ \hline
\multirow{4}{*}{$\eta$=0.5,   $\zeta$=0.0} & Return & 42.80±4.36 & 42.90±4.36 & - & - & - & 42.86±4.32 \\ \cline{2-8} 
 & Sharpe Ratio & 9.81 & 9.85 & - & - & - & 9.93 \\ \cline{2-8} 
 & Quoting Ratio & 0.00+100.00+0.00+0.00 & 0.00+100.00+0.00+0.00 & - & - & - & 0.00+100.00+0.00+0.00 \\ \cline{2-8} 
 & Spread & 3.01±0.78 & 2.87±0.81 & - & - & - & 3.04±0.77 \\ \hline
\multirow{4}{*}{$\eta$=0.1,   $\zeta$ =0.0} & Return & 59.87±4.74 & 60.03±4.07 & - & - & - & 61.57±3.73 \\ \cline{2-8} 
 & Sharpe Ratio & 12.62 & 14.76 & - & - & - & 16.52 \\ \cline{2-8} 
 & Quoting Ratio & 0.00+100.00+0.00+0.00 & 0.00+100.00+0.00+0.00 & - & - & - & 0.00+100.00+0.00+0.00 \\ \cline{2-8} 
 & Spread & 1.49±0.37 & 1.30±0.31 & - & - & - & 1.52±0.32 \\ \hline
\multirow{4}{*}{$\eta$=0.01,   $\zeta$=0.0} & Return & 58.37±4.10 & 56.07±3.92 & - & - & - & 49.95±3.40 \\ \cline{2-8} 
 & Sharpe Ratio & 14.25 & 14.30 & - & - & - & 14.68 \\ \cline{2-8} 
 & Quoting Ratio & 0.00+100.00+0.00+0.00 & 0.00+100.00+0.00+0.00 & - & - & - & 0.00+100.00+0.00+0.00 \\ \cline{2-8} 
 & Spread & 1.60±0.36 & 1.82±0.50 & - & - & - & 1.92±0.51 \\ \hline
\multirow{4}{*}{$\eta$=0.0,   $\zeta$=0.01} & Return & 53.13±3.93 & 53.14±3.89 & - & - & - & 51.03±3.54 \\ \cline{2-8} 
 & Sharpe Ratio & 13.53 & 13.66 & - & - & - & 14.41 \\ \cline{2-8} 
 & Quoting Ratio & 0.00+100.00+0.00+0.00 & 0.00+100.00+0.00+0.00 & - & - & - & 0.00+100.00+0.00+0.00 \\ \cline{2-8} 
 & Spread & 1.89±0.54 & 1.79±0.60 & - & - & - & 2.03±0.33 \\ \hline
\multirow{4}{*}{$\eta$=0.0, $\zeta$=0.001} & Return & 64.61±4.44 & 59.64±4.06 & - & - & - & 62.70±3.88 \\ \cline{2-8} 
 & Sharpe Ratio & 14.56 & 14.70 & - & - & - & 16.15 \\ \cline{2-8} 
 & Quoting Ratio & 0.00+100.00+0.00+0.00 & 0.00+100.00+0.00+0.00 & - & - & - & 0.00+100.00+0.00+0.00 \\ \cline{2-8} 
 & Spread & 1.49±0.27 & 1.53±0.30 & - & - & - & 1.53±0.26 \\ \hline
\end{tabular}%
}
\end{table*}

% \usepackage{multirow}
% \usepackage{graphicx}
\begin{table*}[h]
\centering
\caption{MM with 2-Action Space against Adversarial Training}
\label{MM with 2-Action Space against Adversarial Training}
\resizebox{\textwidth}{!}{%
\begin{tabular}{|c|c|c|c|c|c|c|c|}
\hline
Risk   Coefficient & Desiderata & Fix & Random & A & B & K & All \\ \hline
\multirow{4}{*}{$\eta$=0.0,   $\zeta$ =0.0} & Return & 66.85±11.10 & 61.05±4.13 & 54.52±4.08 & 61.09±4.07 & 54.56±4.08 & 57.73±3.47 \\ \cline{2-8} 
 & Sharpe Ratio & 6.02 & 14.79 & 13.35 & 15.02 & 13.38 & 16.61 \\ \cline{2-8} 
 & Quoting Ratio & 0.00+100.00+0.00+0.00 & 0.00+100.00+0.00+0.00 & 0.00+100.00+0.00+0.00 & 0.00+100.00+0.00+0.00 & 0.00+100.00+0.00+0.00 & 0.00+100.00+0.00+0.00 \\ \cline{2-8} 
 & Spread & 1.69±0.31 & 1.45±0.22 & 1.76±0.30 & 1.44±0.22 & 1.78±0.32 & 1.61±0.26 \\ \hline
\multirow{4}{*}{$\eta$=1.0,   $\zeta$=0.0} & Return & 30.03±3.87 & 32.95±4.21 & - & - & - & 36.07±3.47 \\ \cline{2-8} 
 & Sharpe Ratio & 7.75 & 7.83 & - & - & - & 10.39 \\ \cline{2-8} 
 & Quoting Ratio & 0.01+99.99+0.00+0.00 & 0.00+100.00+0.00+0.00 & - & - & - & 0.00+100.00+0.00+0.00 \\ \cline{2-8} 
 & Spread & 2.29±0.99 & 3.14±0.95 & - & - & - & 2.80±0.55 \\ \hline
\multirow{4}{*}{$\eta$=0.5,   $\zeta$=0.0} & Return & 42.81±4.36 & 42.82±4.33 & - & - & - & 42.13±4.22 \\ \cline{2-8} 
 & Sharpe Ratio & 9.81 & 9.88 & - & - & - & 9.99 \\ \cline{2-8} 
 & Quoting Ratio & 0.00+100.00+0.00+0.00 & 0.00+100.00+0.00+0.00 & - & - & - & 0.00+100.00+0.00+0.00 \\ \cline{2-8} 
 & Spread & 3.01±0.76 & 3.04±0.81 & - & - & - & 2.95±1.04 \\ \hline
\multirow{4}{*}{$\eta$=0.1,   $\zeta$ =0.0} & Return & 59.97±4.75 & 60.07±4.05 & - & - & - & 61.62±3.73 \\ \cline{2-8} 
 & Sharpe Ratio & 12.64 & 14.83 & - & - & - & 16.52 \\ \cline{2-8} 
 & Quoting Ratio & 0.00+100.00+0.00+0.00 & 0.00+100.00+0.00+0.00 & - & - & - & 0.00+100.00+0.00+0.00 \\ \cline{2-8} 
 & Spread & 1.57±0.43 & 1.19±0.24 & - & - & - & 1.48±0.33 \\ \hline
\multirow{4}{*}{$\eta$=0.01,   $\zeta$=0.0} & Return & 58.29±4.06 & 56.01±3.88 & - & - & - & 49.94±3.39 \\ \cline{2-8} 
 & Sharpe Ratio & 14.36 & 14.42 & - & - & - & 14.73 \\ \cline{2-8} 
 & Quoting Ratio & 0.00+100.00+0.00+0.00 & 0.00+100.00+0.00+0.00 & - & - & - & 0.00+100.00+0.00+0.00 \\ \cline{2-8} 
 & Spread & 1.58±0.33 & 1.79±0.49 & - & - & - & 1.93±0.50 \\ \hline
\multirow{4}{*}{$\eta$=0.0,   $\zeta$=0.01} & Return & 56.69±4.17 & 53.10±3.89 & - & - & - & 51.02±3.54 \\ \cline{2-8} 
 & Sharpe Ratio & 13.59 & 13.66 & - & - & - & 14.41 \\ \cline{2-8} 
 & Quoting Ratio & 0.00+100.00+0.00+0.00 & 0.00+100.00+0.00+0.00 & - & - & - & 0.00+100.00+0.00+0.00 \\ \cline{2-8} 
 & Spread & 1.81±0.26 & 1.73±0.51 & - & - & - & 1.99±0.32 \\ \hline
\multirow{4}{*}{$\eta$=0.0, $\zeta$=0.001} & Return & 64.53±4.38 & 59.67±4.05 & - & - & - & 62.69±3.88 \\ \cline{2-8} 
 & Sharpe Ratio & 14.72 & 14.73 & - & - & - & 16.14 \\ \cline{2-8} 
 & Quoting Ratio & 0.00+100.00+0.00+0.00 & 0.00+100.00+0.00+0.00 & - & - & - & 0.00+100.00+0.00+0.00 \\ \cline{2-8} 
 & Spread & 1.42±0.27 & 1.51±0.31 & - & - & - & 1.51±0.26 \\ \hline
\end{tabular}%
}
\end{table*}

% \usepackage{multirow}
% \usepackage{graphicx}
\begin{table*}[h]
\centering
\caption{MM with 4-Action Space against Adversarial Training}
\label{MM with 4-Action Space against Adversarial Training}
\resizebox{\textwidth}{!}{%
\begin{tabular}{|c|c|c|c|c|c|c|c|}
\hline
Risk   Coefficient & Desiderata & Fix & Random & A & B & K & All \\ \hline
\multirow{4}{*}{$\eta$=0.0,   $\zeta$ =0.0} & Return & 65.85±10.85 & 61.03±4.11 & 52.83±3.95 & 60.12±4.00 & 54.51±4.03 & 57.61±3.46 \\ \cline{2-8} 
 & Sharpe Ratio & 6.07 & 14.84 & 13.39 & 15.04 & 13.52 & 16.64 \\ \cline{2-8} 
 & Quoting Ratio & 0.00+98.23+0.29+1.48 & 0.00+99.96+0.00+0.04 & 0.00+99.96+0.00+0.04 & 0.00+100.00+0.00+0.00 & 0.00+99.92+0.00+0.07 & 0.00+99.61+0.19+0.20 \\ \cline{2-8} 
 & Spread & 1.71±0.34 & 1.44±0.23 & 1.73±0.43 & 1.26±0.28 & 1.77±0.29 & 1.56±0.28 \\ \hline
\multirow{4}{*}{$\eta$=1.0,   $\zeta$=0.0} & Return & 30.00±3.87 & 32.94±4.21 & - & - & - & 36.13±3.48 \\ \cline{2-8} 
 & Sharpe Ratio & 7.75 & 7.83 & - & - & - & 10.39 \\ \cline{2-8} 
 & Quoting Ratio & 0.00+100.00+0.00+0.00 & 0.00+100.00+0.00+0.00 & - & - & - & 0.00+100.00+0.00+0.00 \\ \cline{2-8} 
 & Spread & 2.29±1.14 & 2.90±1.11 & - & - & - & 2.84±0.56 \\ \hline
\multirow{4}{*}{$\eta$=0.5,   $\zeta$=0.0} & Return & 42.92±4.29 & 45.66±4.55 & - & - & - & 45.35±4.35 \\ \cline{2-8} 
 & Sharpe Ratio & 10.01 & 10.03 & - & - & - & 10.43 \\ \cline{2-8} 
 & Quoting Ratio & 0.00+96.33+3.67+0.00 & 0.00+99.52+0.00+0.48 & - & - & - & 0.00+97.44+0.00+2.56 \\ \cline{2-8} 
 & Spread & 2.90±0.80 & 1.93±0.71 & - & - & - & 2.07±0.70 \\ \hline
\multirow{4}{*}{$\eta$=0.1,   $\zeta$ =0.0} & Return & 59.95±4.73 & 60.01±4.04 & - & - & - & 61.63±3.73 \\ \cline{2-8} 
 & Sharpe Ratio & 12.66 & 14.84 & - & - & - & 16.51 \\ \cline{2-8} 
 & Quoting Ratio & 0.00+100.00+0.00+0.00 & 0.00+100.00+0.00+0.00 & - & - & - & 0.00+100.00+0.00+0.00 \\ \cline{2-8} 
 & Spread & 1.41±0.31 & 1.21±0.28 & - & - & - & 1.46±0.28 \\ \hline
\multirow{4}{*}{$\eta$=0.01,   $\zeta$=0.0} & Return & 58.25±4.05 & 56.03±3.88 & - & - & - & 51.58±3.48 \\ \cline{2-8} 
 & Sharpe Ratio & 14.40 & 14.43 & - & - & - & 14.84 \\ \cline{2-8} 
 & Quoting Ratio & 0.00+99.47+0.00+0.53 & 0.00+100.00+0.00+0.00 & - & - & - & 0.03+99.20+0.00+0.77 \\ \cline{2-8} 
 & Spread & 1.62±0.34 & 1.76±0.49 & - & - & - & 1.93±0.28 \\ \hline
\multirow{4}{*}{$\eta$=0.0,   $\zeta$=0.01} & Return & 64.72±4.75 & 53.10±3.89 & - & - & - & 51.18±3.54 \\ \cline{2-8} 
 & Sharpe Ratio & 13.63 & 13.66 & - & - & - & 14.46 \\ \cline{2-8} 
 & Quoting Ratio & 0.00+99.87+0.00+0.13 & 0.00+100.00+0.00+0.00 & - & - & - & 0.09+99.11+0.19+0.61 \\ \cline{2-8} 
 & Spread & 1.43±0.26 & 1.66±0.58 & - & - & - & 1.95±0.28 \\ \hline
\multirow{4}{*}{$\eta$=0.0, $\zeta$=0.001} & Return & 64.53±4.36 & 59.64±4.01 & - & - & - & 62.66±3.87 \\ \cline{2-8} 
 & Sharpe Ratio & 14.80 & 14.87 & - & - & - & 16.17 \\ \cline{2-8} 
 & Quoting Ratio & 0.00+100.00+0.00+0.00 & 0.00+100.00+0.00+0.00 & - & - & - & 0.00+100.00+0.00+0.00 \\ \cline{2-8} 
 & Spread & 1.40±0.26 & 1.55±0.24 & - & - & - & 1.46±0.29 \\ \hline
\end{tabular}%
}
\end{table*}

\subsection{Performance of always quoting MMs}
Compared to the performance presented in \cite{spooner2020robust}, the MMs that always quote in this experiment generally have lower average profits in the RN and RA cases, but still maintain high Sharpe ratios due to their lower variance. At the same time, the MMs trained to confront a random adversary exhibit slightly higher Sharpe ratios than those trained against a fixed adversary, but the market makers trained against an all adversary that controls all three parameters simultaneously show the best performance in the evaluation process. However, the MMs trained against adversaries that only control $A$ or only control $K$ perform poorly. Qualitatively, the performance of the MMs who always quote in this experiment is similar to the results presented in \cite{spooner2020robust}. However, due to differences in the programming language and algorithm used, as well as the randomness of the model, quantitatively the results may differ.

\subsection{Performance of MMs able to not quote}
Although the experimental environment and training algorithms for MMs with two and four actions are different from those of always quoting MMs, results exhibit similar patterns, demonstrating the effectiveness of ARL in improving the robustness of market maker strategies.

The experimental results demonstrate that a MM with a rich action space that maintains bilateral quotes during trading has essentially the same value of the Sharpe ratio as an always-quoting MM, mainly because the latter is used by the former when it decides to submit two-sided quotes. In some cases, if a MM chooses to mitigate market risk by not quoting or submitting one-sided quotes, its Sharpe ratio is slightly higher than the Sharpe ratio of the corresponding always-quoting MM. However, on the whole, the results of this experiment show that even with a rich action space, MMs tend to make two-sided quotes to obtain more bid-ask spreads, which is consistent with the essence of providing market liquidity. The average quoting ratios in the experiments exceeded 95$\%$, thus meeting basically all the aforementioned market rules. 

This may be because the market makers' quoting strategy trained by ARL is already able to control risks through reasonable quotes. Moreover, even if the market fluctuates, the risk brought by one unit of inventory is limited. The model assumes that at each time step if the agent matches an ask or bid order, it can only trade one unit, which may make the agent less sensitive to inventory risk. In other words, the inventory risk is relatively small, so the market maker may be more inclined to take risks and continue quoting to obtain bid-ask spread even if the market fluctuates. For example, MMs with two actions under the RN situation, have two choices at each time step. If they choose to quote, they will call the stored always-quoting MM strategy to give a two-sided quote. If they successfully match orders, such as matching both bid and ask orders, only matching bid orders, or only matching ask orders, the agent will receive a quote spread as a trading reward, and then update the current inventory value with the new market price in the next time step. If they choose not to quote, they do not need to quote or match orders and will directly update the current inventory value with the new market price in the next time step. Therefore, in a relatively stable market with low exposure to inventory risk, market makers will tend to continue quoting to obtain more income from the  bid-ask spread.

\section{Conclusions} 
In this paper, we proposed a flexible market maker quoting strategy that is robust in different adversarial environments. We designed a market maker that can evaluate the market environment and give reasonable spreads based on the inventory at times when it is expected to make profits. In general, the market environment is complex and dynamic, and keeping quoting regardless of the situation may trigger losses for the market maker. Therefore, it is necessary for market makers to make use of quotation rules and review the situation to actively and adaptively decide when to submit two-sided quotes, but to ensure that they meet certain quoting ratio requirements as qualified liquidity providers. To this end, our proposed market maker's strategy is shown to ensure the provision of market liquidity, while hedging potential risks by not always quoting. Our results demonstrate that a flexible quoting mechanism helps market makers dynamically manage the inventory risk, while profiting from the difference in the bid-ask spread.

For future work, it would be interesting to explore more sophisticated models that simulate a more realistic market environment. Firstly, we can improve the current quotation mechanism by setting a stricter quotation ratio limit. It is important to ensure that the market makers have the opportunity to hedge their risks, while providing as much liquidity as possible to the market. Secondly, it is necessary to improve the configuration of the training network -- for example, by increasing the number of training episodes, and thus reducing the randomness of the experiments. We could also employ other reinforcement learning algorithms and compare them with the performance of the currently implemented. In addition, it is interesting to explore richer state spaces for the market makers, that currently only include time and inventory, by adding more features, such as charging transaction costs or giving rebates to the market makers. Furthermore, this paper mainly focuses on inventory risk. However, if market makers were allowed to monitor market volatility in a more volatile market simulation environment and consider it as an indicator for measuring adverse selection risk, it might provide them with better tools to comprehensively manage risk and make quoting decisions. Finally, the action space for the market makers could be further expanded. Currently, they are only required to give a bid/ask spread if they are willing to quote, but the default volume is 1 unit. It would, however, be sensible for the market makers to be able to quote the volume in which they are willing to trade.

%%\section{Appendices}
%%\section{Acknowledgments}

\bibliographystyle{ACM-Reference-Format}
\bibliography{main}

\end{document}